\documentclass{PoS}
\usepackage{journals, amsmath}
\usepackage{cite}

\title{Where are the Ultra High Energy Cosmic Ray (UHECR) originated?}

\ShortTitle{Where are the UHECR originated?}

\author{\speaker{Daniele Fargion},$^{a,b}$ {{Pier Giorgio} {De Sanctis Lucentini}}$^{~c}$ and Maxim Yu. Khlopov$^{~d,e,f}$\\
\llap{$^a$}  Physics Department \& INFN Rome1, Rome University 1, P.le A. Moro 2, 00185, Rome, Italy\\
\llap{$^b$}  MIFP, Via Appia Nuova 31, 00040 Marino (Rome), Italy\\
\llap{$^c$}Physics Department, Gubkin Russian State University (National Research
      University),\\
      65 Leninsky Prospekt, Moscow, 119991, Russian Federation\\
\llap{$^d$} National Research Nuclear University "MEPHI" (Moscow Engineering Physics Institute),  Kashirskoe Sh. 31, Moscow 115409,  Russia\\
\llap{$^e$} Centre for Cosmoparticle Physics "Cosmion" 115409 Moscow, Russia\\
\llap{$^f$} APC laboratory 10, rue Alice Domon et Lonie Duquet 75205 Paris Cedex 13, France\\

E-mail: \email{daniele.fargion@roma1.infn.it}, \email{desanctislucentini.pg@gubkin.ru},
\email{khlopov@apc.in2p3.fr}
}

\FullConference{XII Multifrequency Behaviour of High Energy Cosmic Sources Workshop\\
                12-17 June, 2017\\
                Palermo, Italy}

\abstract{
We consider the recent results  on UHECR (Ultra High Energy Cosmic Ray)  composition and their distribution in the sky from ten EeV energy (the dipole anisotropy) up to the highest UHECR energies and their clustering maps: UHECR  have been found mostly made by light and lightest nuclei. We  summarized the arguments that favor a few localized nearby extragalactic sources for most UHECR as CenA, NG 253, M82.
 We comment also on the possible partial role of a few remarkable  galactic UHECR sources.
Finally we revive the eventual role of a relic neutrino eV mass in dark hot halo (hit by ZeV neutrinos) to explain the new UHECR clustering  events centered around a very far cosmic AGN sources as 3C~454.

}

\begin{document}

\section{Introduction}\label{sec:intro}

The Ultra High Energy Cosmic Rays (UHECR) seem to be arrived at their final understanding stages. Because of their nature, the hadron charged Cosmic Rays (CR) are bent and smeared by magnetic fields both from near and far sources. The lepton CR, instead, lose energy very fast and have a local origin, bounded at lowest energy (TeVs). However, the highest energetic CR (EeV or ZeV) should be able to point back to their source with a good directionality thanks to their rigidity.
Moreover the UHECR disruption and fragmentation (due to the photo pion or the photo dissociation caused by the cosmic radiation interaction) make the UHECR confined into a local Universe. These distances are respectively 40 Mpc for UHECR nucleon or 4 Mpc light nuclei due to the GZK cutoff, the interaction with the cosmic microwave background photons. 
The heaviest CR like iron are too much bent by their large charges to be disentangled in the sky. No iron-like UHECR astronomy at horizon. Proton, the most popular courier must fly nearly straight (clustering within a few  degrees,$\simeq 3^{\circ}$ at $10^{20}$ eV).
But there are not such characteristic narrow UHECR clusterings. Lightest UHECR nuclei (He,D,Li,Be..) directions are more smeared, but still quite collimated, up to ten or twenty degrees, as wide as the first observed UHECR anisotropy; lightest UHECR nuclei (He like) are constrained in a few Mpc Universe, playing a key role in our UHECR understanding.
The earliest $3 \cdot 10^{20}$ eV energetic UHECR events recorded by Fly's Eye and the abundant events over GZK cutoff  by AGASA were apparently overcoming any energy cutoff: they drove in 1997 to the Z Boson resonant model based on the UHE ZeV neutrino (from a far cosmic AGN) scattering on the relic cosmic ones (with eV mass).

 More  UHECR events and statistics recorded by Hires and by Auger array detectors had shown, on the 2000s, the expected UHECR GZK-like cutoff. Later on 2007 a first Auger map of $26$ UHECR events led most authors to claim a first correlation with the Super Galactic (SG) local plane (located as foreseen in a GZK volume).
 However, a few years later, with the additional UHECR data the loss of any UHECR SG correlation drove to an undistinguished  cosmic source location; few clustered multiplets of UHECR events at twenty EeV, around Cen~A had been foreseen and observed in the triennium, 2009-2011.

 The more recent results, with unprecedented statistics, of the slant depth of the shower maximum, $X_{max}$, in the years of 2013 -- 2015, from Auger and Telescope Array (TA) ultimately convinced most authors that UHECR above 10 EeV are mainly light and lightest nuclei.

 The large unexpected anisotropy ($7\%$) of ten EeV UHECR energy have shown in 2017 a remarkable (5 sigma) signal showing an amazing UHECR dipole anisotropy: neither the galactic center nor the nearby Super Galactic Plane did show such a signal. We show a different mixed galactic and extragalactic possible sources.

  In the beginning of 2018 the updated 39 EeV and 60 EeV UHECR event maps by Auger are showing enhanced clusterings on the nearest AGN or starburst sources localized in the local universe, as it has been earlier suggested (2008-2015).  We show these chains of arguments that led to identify first extragalactic Cen A, M82, NGC~253 and Fornax~D as the main nearby sources. We explain as the dipole anisotropy might find a main component (at 10 EeV UHECR) born within nearby galactic sources. We finally underline two uncorrelated clustering around 3C~454 and Markarian~421: we advocate for them a ZeV (UHE) neutrino model (by cosmic far AGN) hitting onto a cosmic relic neutrino (with an eV mass, maybe sterile one) making a Z boson resonance overcoming the GZK cutoff.
\subsection{A final UHECR astronomy? The article structure}
The discovery of Cosmic Rays (CR) in the last century motivated many experiment groups to accumulate, with great efforts, steady and large data records aiming at determining its charged nuclear nature, its spectra and composition.
The (mysterious) absence of any cosmic magnetic monopoles reflects into the presence of large scale (planetary, stellar, galactic, cosmic) magnetic fields.
The charged nuclei or nucleon in CR are therefore easily deflected by the astrophysical magnetic fields via Lorentz forces.
Therefore, CR are loosing their primordial  directionality.

However, the CR of the highest energies above $10^{19}$ eV (tens EeV), if protons, suffer less and less from deflection promising a geometrical connection with their original sources.
The Ultra High Energy Cosmic Ray (UHECR) neutrons are neutral and un-deflected but bounded by instability within a very narrow cosmic radius: $\simeq Mpc \frac{E_{n}}{10^{20} eV}$.
In last two decades, largest area detectors for UHECR as the AGASA \cite{AGASA(1998)}, the  High Resolution Fly's Eye (HiRes), the Telescope Array (TA), the Pierre Auger Observatory (Auger) were able to record several hundreds of $6\cdot10^{19}$ eV UHECR, near or above the GZK cutoff \cite{GZK(1966),GZK(1966)b}.
Here we tried to briefly review the last decades of road map toward the disentanglement of the most probable UHECR sources.\\

In the first section we discuss the remarkable absence of Virgo in the UHECR clustering since 2007 \cite{PAO(2007)} up to 2015, in contrast with the appearance of the so called \emph{two hot spot} UHECR anisotropy (see Fig.\ref{Fargionfig1}, the most dense source masses in a GZK allowed volumes).\\

In the second section we summarize the composition signature of these UHECR, whose nature has been widely assumed
up to few years ago (2016) to be an hybrid mixture of iron and proton.
Today it is quite openly agreed that they are light nuclei (or lightest ones) because of their average depth of the shower maximum~\cite{PAO(2017b)}.\\

In the third section we remind the characteristic bending (or deflection) angle expected for random or coherent deflection by Lorentz forces \cite{Fargion(2009b)}, mainly due to galactic magnetic fields. We show how these lightest nuclei UHECR bendings are well compatible and tuned with the observed anisotropy, while the proton or iron ones are too small or too wide, as it has been also recently recognized \cite{PAO(2017b)} (see note page 26, References 83-87).\\

In the fourth section we address the very recent dipole anisotropy found by Auger at 10 EeV with an unexpectedly large ($7\%$)
weight \cite{PAO(2017)}. Its location in the sky and its statistical relevance suggest  a combined, but ruling galactic origin (Vela, Crab, LMC, SMC) of these UHECR with a small extra galactic presence.\\

On the fifth section we discuss and update the foreseen lightest nuclei fragmentation from Cen~A at twenty EeV \cite{Fargion(2009b)}  and the later somehow observed UHECR multiplets at 20 EeV, possibly because of the He UHECR fragments spread in a  vertical clustering (in a plane nearly orthogonal to the galactic plane) as well as a similar LMC and SMC possible UHECR clustering multiplet published in an (often neglected) article \cite{PAO(2011)}.\\

On the sixth section we show the most energetic $60$ EeV and most narrow UHECR clustering maps and their most probable sources.
In the same section we also discuss the case of an unexplained clustering correlated to a half cosmic radius\footnote{half cosmic radius, or  $0.859 $ redshift or at a time flight distance of $7.7$ Gly or $2.5$ Gpc} far source: 3C454.
We proposed \cite{Fargion.2018} and confirm here that this clustering, if growing (as it is) and being confirmed, might be associated and originated by a far beamed AGN, whose ZeV neutrino are scattering onto a relic cosmic neutrino background (with $\simeq 1-2$eV mass) via Z boson resonance (the so called Z burst model) \cite{Fargion.et.al.(1999), Fargion2007Splitting, Fargion2004Colaiuda, Fargion2017NarrowClustering}.
Indeed the UHE ZeV neutrinos, emitted by far AGN do not interact with the cosmic photons and do not suffer of a GZK cutoff.
A ZeV UHE neutrino does scatter well with relic neutrinos in Z boson resonant peaked cross section \cite{Fargion.et.al.(1999)}.
The UHE  Z boson decay in pions, nucleons and anti nucleons might be the observed UHECR and its hard gamma brightening
 \cite{Fargion2004Colaiuda,Fargion2017NarrowClustering}.
 These relic neutrinos with a mass in a cosmic presence of a relic hot universe, might be also tuned with the exciting (but still controversial) sterile neutrino existence (claimed with a mass well tuned at $1.6$ eV). Similar arguments apply also to  a much nearer (but still too much far respect to the GZK distances) AGN Mrk~421, (at redshift: z=0,0308 or at distances ranging from 122 Mpc to 133 Mpc), shining around the north galactic Hot Spot found by TA.\\

 Finally in the conclusion section  we foresee future signals in UHECR as well in their photo-dissociations neutrinos;
 we remind that lightest nuclei UHECR lower energetic neutrino fragments might explain better the more and more evidences of an absence of a GZK \cite{GZK(1966),GZK(1966)b} EeV neutrinos secondaries (expected  but unobserved by the still popular UHECR proton candidates).
 Lightest UHECR fragments provide a source for lower energy neutrino signals (at tens PeV) at smaller fluency,
 better observable in spectacular upward tau air-shower signals \cite{Fargion(2000), Fargion.et.al.(2004), PAO(2009), PAO(2008)} and maybe in future IceCube records.
This occur because lightest UHECR nuclei suffer at lower energy (EeVs) from a photo-nuclear-dissociation
by nuclear resonance already present at energies (and survival distances) below the one needed for the photo-pion production, leading to the GZK cutoff (as discussed below, see e.g. the Fig. \ref{Fargionfig5})."

\section{The Virgo Absence and the two Hot Spots}

\begin{figure}[t]
\begin{center}
\includegraphics[width=0.98\textwidth]{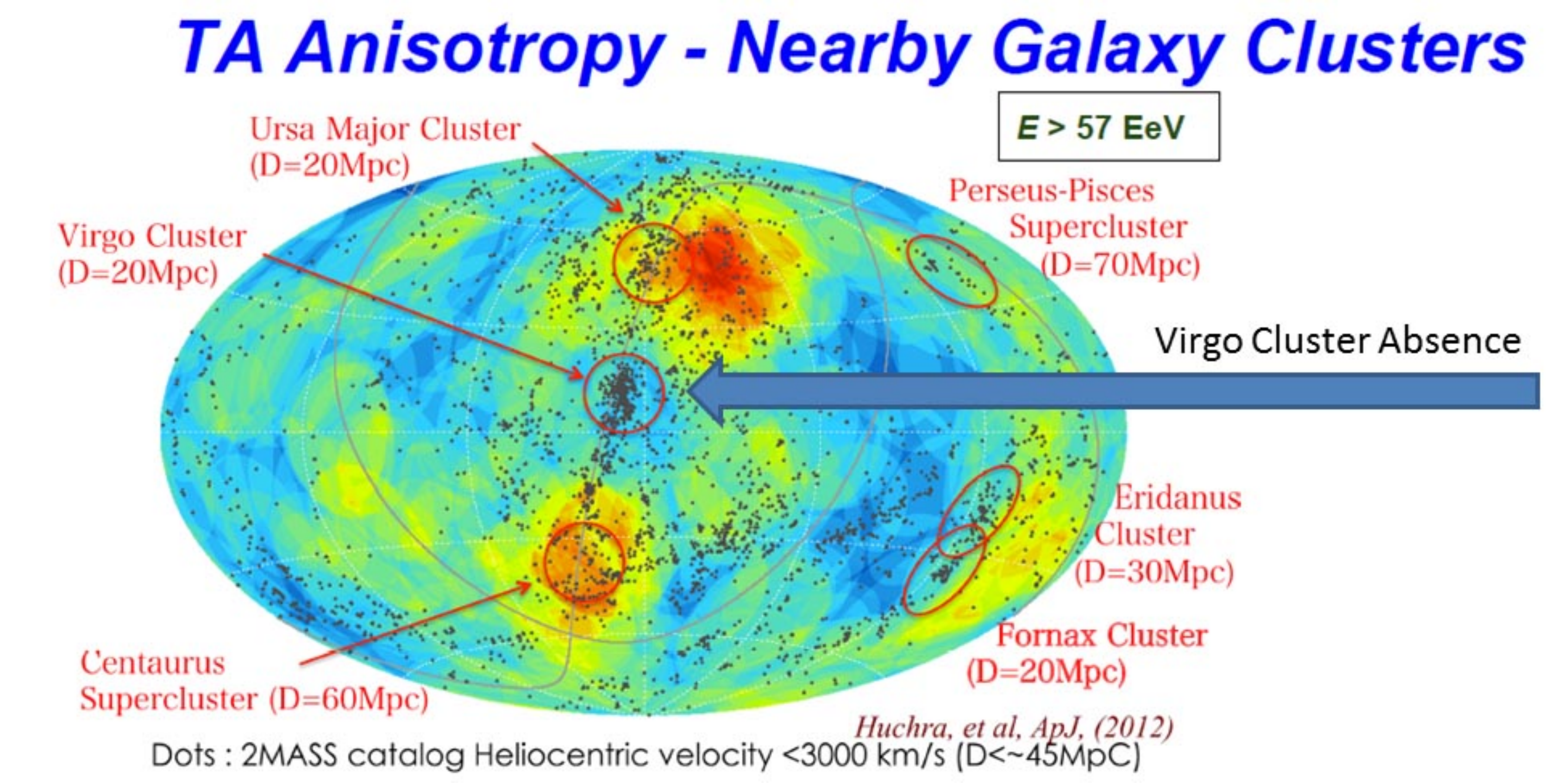}
\caption{The combined  map, in equatorial coordinates, based on the 2015 Auger and TA presentations where the two red hot spot UHECR clustering signals, North for TA and South for Auger,  are showing the main UHECR dense event rate. The dark dots are the main galaxies in nearby GZK volumes\cite{Huchra2012RedShift}. This UHECR map is missing, at its center, of the most abundant and expected Virgo galaxy infrared cluster sources, the heaviest mass concentration within a GZK volume. To explain the absence of the UHECR events from Virgo  we imagined and suggested a filter. We claimed therefore that UHECR were mostly lightest nuclei that are too fragile and cannot reach us above a few Mpc inside the cosmic big bang thermal bath. Indeed Virgo lay at 20 Mpc while the lightest nuclei interaction distance it is below a few Mpc.
Only a decade later the light nuclei UHECR have been accepted by Auger  collaboration team (see note at page 26 of \cite{PAO(2017b)}) because of the observed UHECR average depths of the shower maximum.}
\label{Fargionfig1}
\end{center}
\end{figure}
A picture is worth a thousand words. Indeed in the Fig.\ref{Fargionfig1} here below, one may notice that the crowded black dots galaxies in the map center, where Virgo cluster lay, all of them within the GZK volumes (volume allowed for proton-cosmic photon $2.75$~K opacity): these dots are mainly crowded at the center of the equatorial  map.
However, the two earliest and main red colored UHECR Hot Spots were observed respectively by TA (2014-2016, North sky) and by Auger (2007-2018, South sky) are not correlated at all with the Virgo central direction.
This remarkable Virgo absence is the Rosetta stone that forced us \cite{Fargion(2008)}, a decade ago, toward the lightest\cite{Fargion(2009)} UHECR nuclei nature. Indeed we suggested the presence of a severe filter to stop the expected Virgo abundant UHECR events; if UHECR were protons the center of this map in the Fig.\ref{Fargionfig1} should shine brightly. The unique filter we found  it is the UHECR composition. Indeed  the lightest UHECR might be confined at distances well below the Virgo (20 Mpc) ones, see following Fig.\ref{Fargionfig5}.
This occurs because the lightest UHECR nuclei suffer soon of a photo-nuclear-dissociation by nuclear resonance at energies lower than the one required for photo-pion production (leading to GZK cutoff) and distances (as discussed below, see following Fig.\ref{Fargionfig5}).
The Virgo UHECR event absence  was (and still remains) the first and main Auger cosmic lesson recorded from more than a decade.

\begin{figure}[t]
\begin{center}
\includegraphics[width=0.98\textwidth]{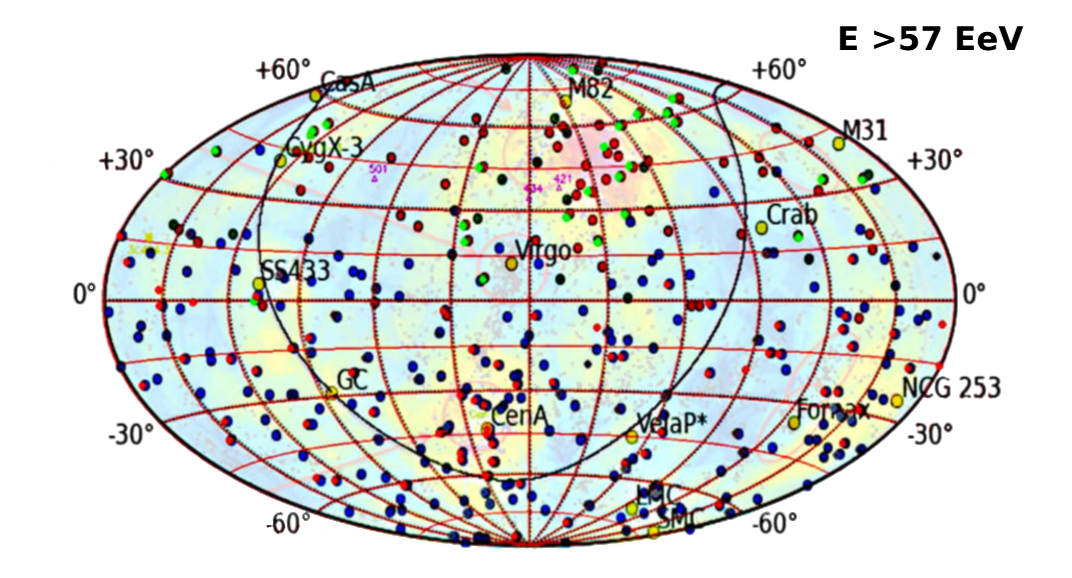}
\caption{As the previous map, here we show the same celestial coordinates map for UHECR
with underline hot spot clustering and individual events (by Auger,  blue dots, TA, red and green dots) whose data have been published \cite{Fargion2015Meaning}. In addition one has the name label of several relevant sources that will be discussed in detail in last section. }
\label{Fargionfig2}
\end{center}
\end{figure}

\begin{figure}[t]
\begin{center}
\includegraphics[width=0.98\textwidth]{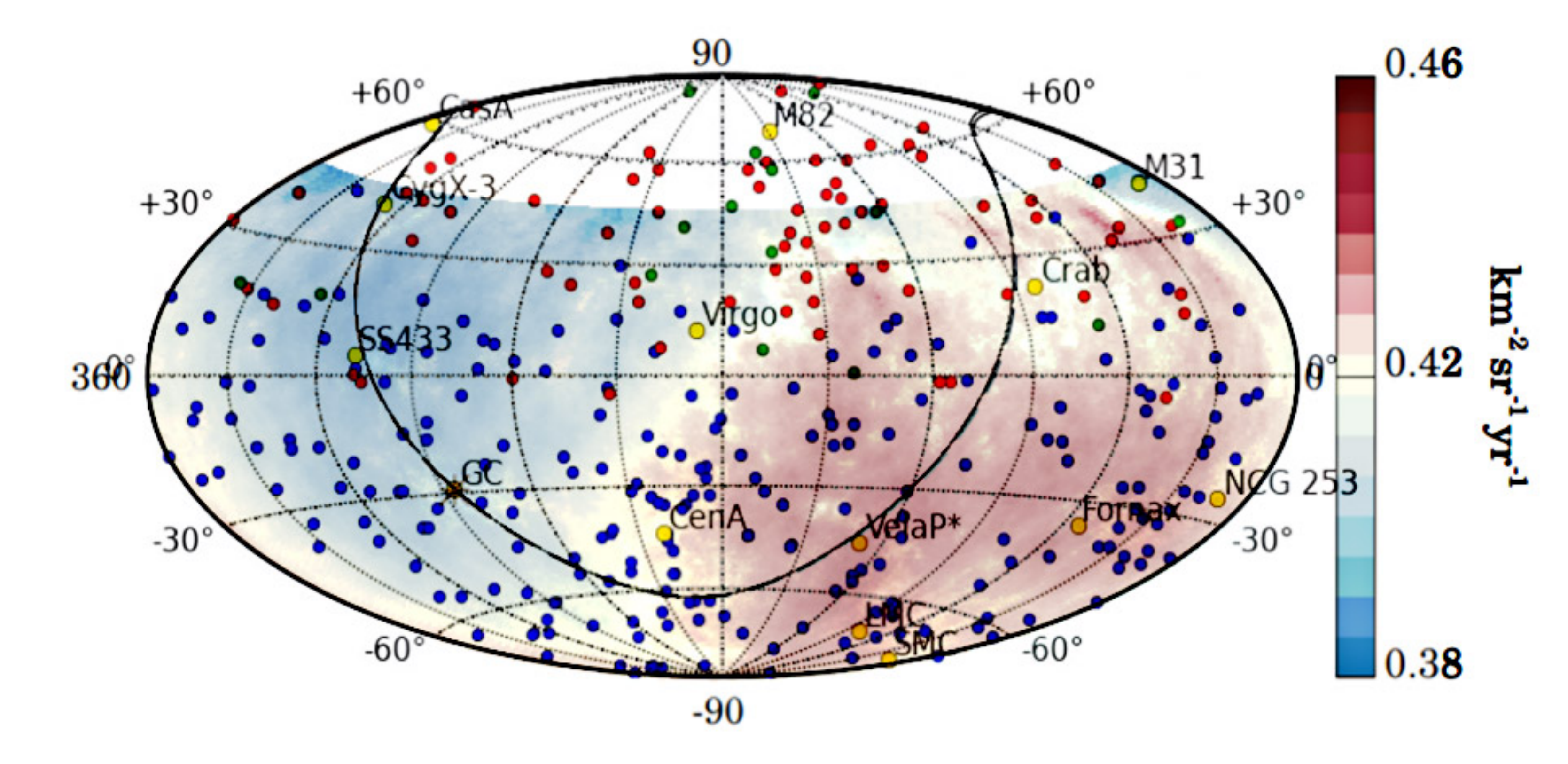}
\caption{The very recent dipole anisotropy shadow found at $8\cdot 10^{18}$ eV by Auger over the other UHECR at GZK energies,  $6\cdot 10^{19}$ (by Auger,  blue dots, TA, red and green dots) on last decade \cite{Fargion2015Meaning}. The presence in the shadow area of well active galactic sources as Vela (the brightest gamma source in Fermi GeV sky), LMC, SMC, Crab and NGC~253, Fornax~D, might be the origin of a very local  anti-galactic center emission.}
\label{Fargionfig3}
\end{center}
\end{figure}

\begin{figure}[t]
\begin{center}
\includegraphics[width=0.98\textwidth]{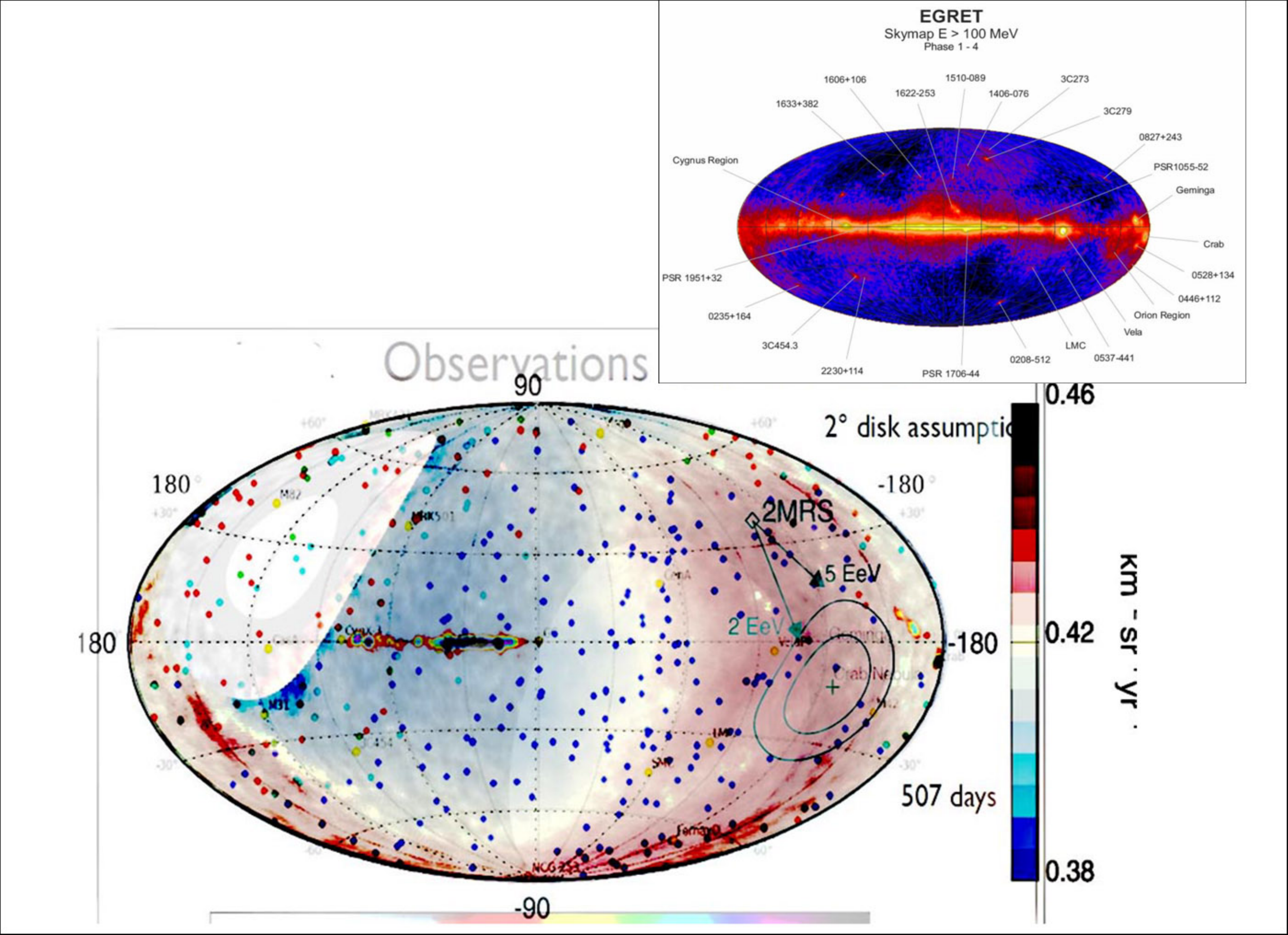}
\caption{As above, but in galactic coordinates, the very recent dipole anisotropy shadow found about ten EeV (and $5\cdot 10^{18}$,$2\cdot 10^{18}$ eV dipole centers)  by Auger over the highest UHECR at GZK energies, $6\cdot 10^{19}$ (by Auger, blue dots, TA, red and green dots) in the last decade \cite{Fargion2015Meaning}. Underlined is also the TeV HAWC (High-Altitude Water Cherenkov Observatory) map. The presence in the shadow area of well active galactic sources as Vela (the brightest galactic gamma source in Fermi GeV sky, see on the top Egret gamma map), LMC, SMC, Crab and NGC~253, Fornax~D, might be the origin of this very local  anti-galactic center emission.}
\label{Fargionfig4}
\end{center}
\end{figure}

\section{The UHECR Lightest Nuclei smeared imprint}
Since the 2007 Auger article there were three additional, not secondary messages, requesting a lightest UHECR nature:
the Cen~A unique clustering and its angular size spread (nearly ten degree size); the Cen~A distance ($ \simeq 3.8 Mpc$) is well consistent with UHECR lightest nuclei survival distance,  see Fig. \ref{Fargionfig5}; Cen~A is the brightest radio, nearest and brightest gamma AGN candidate source.  Both messages (the Cen~A  few Mpc distance and its angular deviation) are consistent with lightest nuclei deflection.
The third hint for the UHECR lightest nuclei nature was derived by Auger the deviation of UHECR average depth of the shower maximum ($X_{max}$ for short) from a proton like toward a light nuclei model signature.
Therefore, while since 2007 the main stream of models for UHECR was preferring an hybrid  iron and proton composition, we preferred UHECR light and lightest nuclei interpretation \cite{Fargion(2008)}, \cite{Fargion(2009)}.
It took nearly a decade to reach  a consensus on the main light and lightest  UHECR nuclei composition \cite{PAO(2017b)}, see Fig.\ref{Fargionfig5}.

\section{Lightest UHECR deflection  in random or coherent path walk and their delay}
The magnetic fields are able to deform the charged CR trajectory in an almost predictable way.
For UHECR with energies above few tens EeV, the deflection occurs from far cosmic distances and by galactic near fields.
The galactic magnetic field is mainly correlated with the galactic arms. The flight across
our galactic plane must suffer of a random bending (or deflection) (if the trajectory it is skimming on our galactic planes, as for the case of Cen A sources) \cite{Fargion(2009b)}.
The galactic field plays a minor role if the path of UHECR is orthogonal to the galactic plane as for the North or South galactic Pole directions.
In this case the magnetic fields are not alternate as along the galactic arms on the galactic plane. Therefore the deflection may be simple a coherent one, not a random one \cite{Fargion(2009b)}.

This may be the case of the AGN  M82 UHECR trajectories that are coming from the pole of our galaxy.
Here we may consider the most general random walk bending (or deflection). While the random walk orthogonal to the galactic plane may lead to an \textit{up-down} bending, orthogonal to the galactic plane disk, as for Cen A, the coherent bending might be the cause of an asymmetric disposal of the UHECR cluster of events with respect to the optical or gamma source candidate.
 In  case of UHECR lightest nuclei, the deflection angle is scaling with the atomic number Z, respectively
($Z_{\mathrm{He}}=2$, $Z_{\mathrm{Li}}=3$, $Z_{\mathrm{Be}}=4$, $Z_{\mathrm{B}}=6$)
because of the extragalactic and mainly because of the galactic deflection angle. Let us evaluate therefore both the two main contributions:
\begin{multline}\label{anglegal}
\alpha^{gal}_{\mathrm{He}}\simeq\\
15.5^\circ\left(\frac{Z}{Z_{\mathrm{He}}}\right)\left(\frac{E}{6\cdot10^{19}\,\mathrm{eV}}\right)^{-1}\left(\frac{D}{20\,\mathrm{kpc}}\right)^{1/2}\left(\frac{d_c}{\mathrm{kpc}}\right)^{1/2}\left(\frac{B}{3\,\mu\mathrm{G}}\right)
\end{multline}
\begin{multline}\label{angleex}
\alpha^{ex}_{\mathrm{He}}\simeq\\
3.28^\circ\left(\frac{Z}{Z_{\mathrm{He}}}\right)\left(\frac{E}{6\cdot10^{19}\,\mathrm{eV}}\right)^{-1}\left(\frac{D}{4\,\mathrm{Mpc}}\right)^{1/2}\left(\frac{d_c}{\mathrm{Mpc}}\right)^{1/2}\left(\frac{B}{\mathrm{nG}}\right)
\end{multline}
which leads to a total of $$\alpha^{ex}_{\mathrm{He}}+\alpha^{gal}_{\mathrm{He}}\simeq18.7^\circ$$ in good agreement with the observed Hot Spot spread angle size. The delay time is mostly due to the extra galactic magnetic field; thus the time delay between UHECR after its photon direct one, can be evaluated as:
\begin{multline}\label{time2}
\Delta\tau_{\mathrm{He}}\simeq\\
2666\left(\frac{Z}{Z_{\mathrm{He}}}\right)^2\left(\frac{E}{6\cdot10^{19}\,\mathrm{eV}}\right)^{-2}\left(\frac{D}{4\,\mathrm{Mpc}}\right)^{2}\left(\frac{d_c}{\mathrm{Mpc}}\right)\left(\frac{B}{1\,\mathrm{nG}}\right)^2\,\mathrm{yr}
\end{multline}

Such a short cosmic delay allows a nominal  Cen~A (as well as the nearest M82, NGC~253) AGN or starburst galaxies to shine
UHECR while being still active in optical, gamma, or X-ray, or radio band, contrary to the alternative proposal of much distant AGN \cite{Kampert(2016)}, \cite{Deligny(2017)} sources, whose charged nuclei random walk have a time of flight (from a hundred Mpc distances) that should suffer from a huge (million years) delay respect to the direct gamma photons; therefore, most far UHECR are very possibly uncorrelated \cite{Fargion.2018} with their present gamma observations \cite{Joshi(2018)}.
The main message in conclusion is that the lightest nuclei as Helium might be the first and the main responsible courier of UHECR, source of the hot spot clustering observed by TA and Auger, in the last decade.
The UHECR source candidates and their coordinates will be described in the final sections. In the Fig~\ref{Fargionfig2} map  we updated the two spot clusterings in celestial coordinates with their underlined (published in detail, 2015) UHECR event map.
The map will help us to better and better disentangle the UHECR sources \cite{2015arXiv151208794F}, adding also the last (updated but averaged, 2018) anisotropy maps \cite{PAO(2017)} in the last section.


\section{The UHECR 10 EeV dipole anisotropy in celestial and galactic maps}
We have to  consider now the quite recent dipole UHECR map at ten EeV energy.
Note in the Figs. \ref{Fargionfig3} and \ref{Fargionfig4} the shadowed area, where the UHECR exceed the background, it is not pointing to our galactic center, nor to nearest cosmic denser sky (as Virgo cluster or the great attractor one). The dipole is pointing versus the anti-galactic center where the loud Crab is located and where the galactic Vela, LMC, SMC sources (as well as the nearby Fornax~D, NGC~253) are also shining. Therefore, the latter sources may be mainly the galactic ones, offering an explanation  of this  remarkable high UHECR over-abundance, as discussed in the final section.
Indeed, the other dipole adimensional anisotropy, as the cosmic black body one it is much smaller ($2  \cdot 10^{-3}$).
We claimed \cite{Fargion.2018} and confirm  that there are not any known extragalactic candidates able to fit the dipole anisotropy \cite{PAO(2017)}; the galactic origin candidacy, made by the above few sources, is probably the best one also with an AGN NGC~253 extragalactic component.
In particular, the random bending (or deflection) at 10 EeV for an Helium like nuclei is spread in a reasonable agreement with the observed dipole size $\pm 90^{o}$:
\begin{multline}\label{DIPOLE}
\alpha^{DIPOLE}_{\mathrm{He}}\simeq
93.0^\circ\left(\frac{Z}{Z_{\mathrm{He}}}\right)\left(\frac{E}{10^{19}\,\mathrm{eV}}\right)^{-1}\left(\frac{D}{20\,\mathrm{kpc}}\right)^{1/2}\left(\frac{d_c}{\mathrm{kpc}}\right)^{1/2}\left(\frac{B}{3\,\mu\mathrm{G}}\right)
\end{multline}

\begin{figure}[ht!]
\centering
  \begin{minipage}{1\textwidth}
  \centering
    \includegraphics[width=0.90\textwidth]{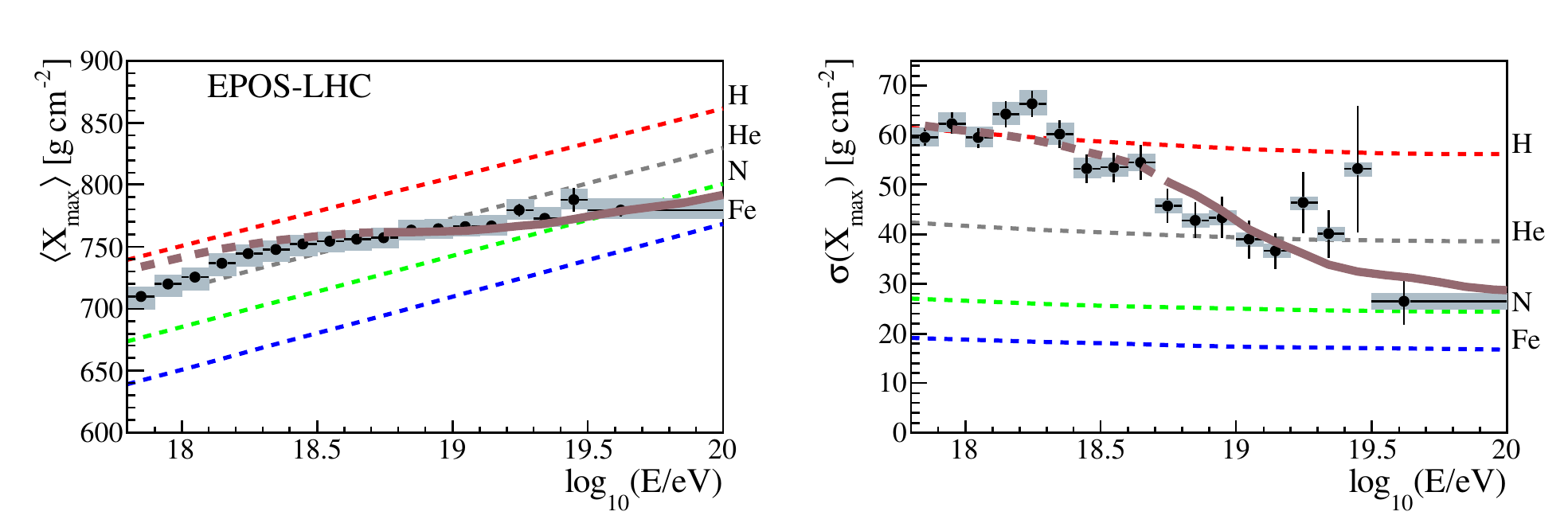}
    \begin{minipage}{.5\textwidth}
      \centering
      \includegraphics[width=0.90\textwidth]{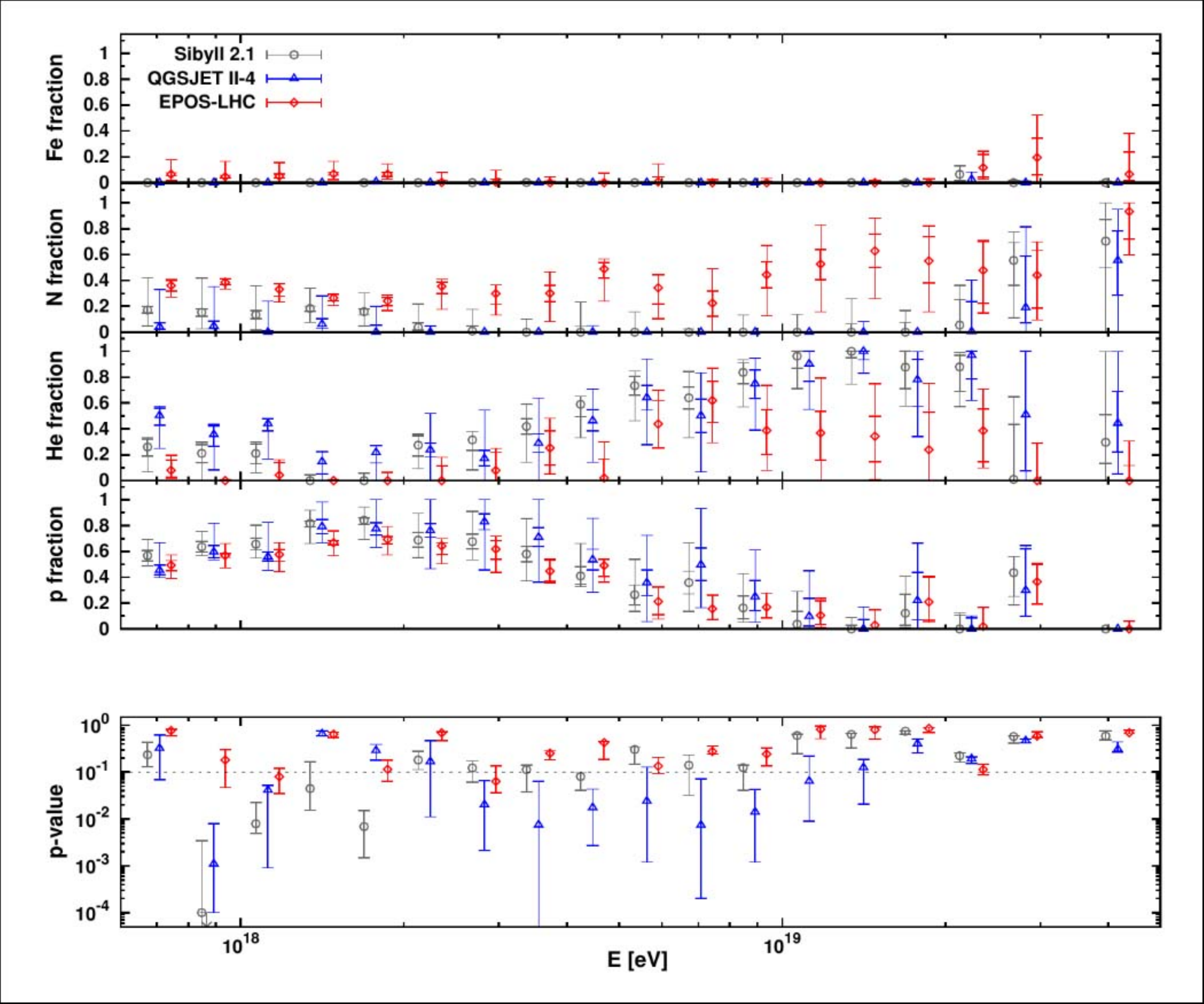}
    \end{minipage}%
    \begin{minipage}{.5\textwidth}
      \centering
      \includegraphics[width=0.90\textwidth]{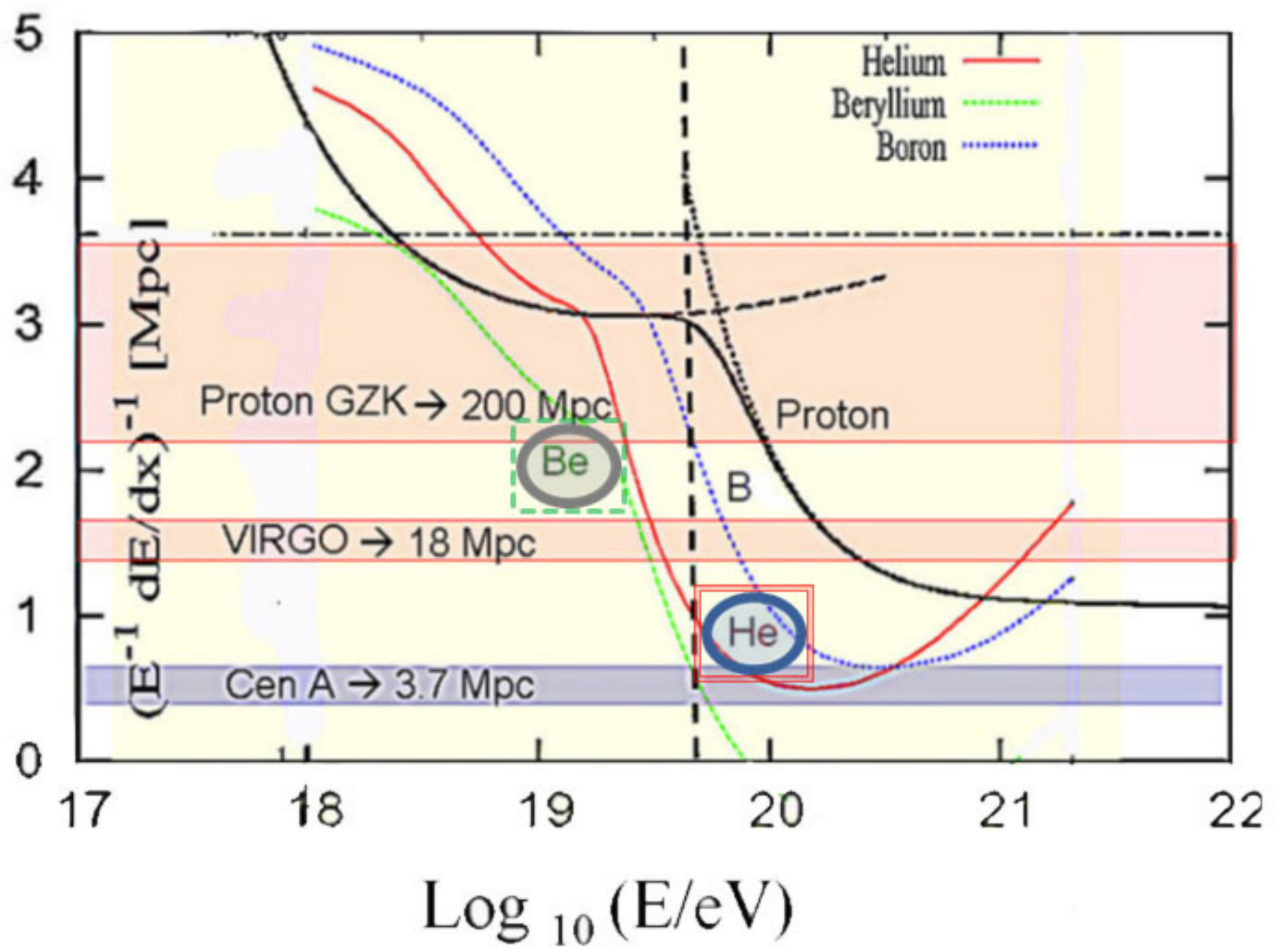}
    \end{minipage}
 \end{minipage}
  \caption{{\bf Top:}
  The average and the standard deviation for the $X_{\max}$ distribution (maximum depth of the shower longitudinal development as measured by the fluorescence detector), as predicted assuming EPOS-LHC UHECR-air interactions for the  model (brown) versus pure $^1$H (red), $^4$He (grey), $^{14}$N (green) and $^{56}$Fe (blue), dashed lines. Only the energy range where the  brown lines are solid is included in the fit, as described in \cite{PAO(2017b)}. The main light and lightest nuclei (He, N) ruling fit above $10^{19}$ eV energy it is obvious.
  {\bf Bottom Left:} a four component fit of the slant depth $X_{max}$  distribution. Upper panels show the fitted fraction of $p$, $He$, $N$ and $Fe$ nuclei. The lower panel shows the quality of the fit. The results from individual hadronic interaction models were slightly shifted in energy for better viewing (Sibyll 2.1 to the left, EPOS-LHC to the right)\cite{PAO(2017c)}. Once again the dominance of the light and lightest UHECR composition (versus proton and iron paucity) above ten EeV it is quite well tuned.
  {\bf Bottom Right:} The consequent allowed distance due to nuclear-photo-dissociation for lightest UHECR nuclei and due to photo-pion production for proton.
  {The interaction lengths of lightest UHECR nuclei by photo-nuclear dissociation it is mainly due to the giant nuclear resonance.
  The UHECR light nuclei survival distance is shown as a function of their energy. The lightest nuclei (colored curves) UHECR GZK distances and  their cutoff occur earlier than for proton photo pion destruction, shown in figure as a black curve.
  The lightest nuclei as He, Li, Be, B are the most favorite ones; Li, Be, B may well mimic the nitrogen N light nuclei behavior: these lightest ones all are banned to arrive from Virgo Cluster, giving a clear answer to the puzzling Virgo UHECR absent signals. The Nitrogen is not banned from Virgo, but it is also widely spread in the sky.
  The three nearest extragalactic AGN or starbursts, Cen A, M 82, NGC~253, are all within the  lightest UHECR nuclei survival distances. Therefore, they could (as they do)  contribute to the observed main UHECR clustering anisotropy observed in recent decade, as shown above  also in celestial and galactic coordinate.}
  }
  \label{Fargionfig5}
\end{figure}

Moreover, the Auger mass composition measurements at 10 EeV favour a dominant He-like (lightest) nuclei, a partial light ones (possibly N, but also Be, B may fit)  and the near absence of proton or iron nuclei,  see Fig.~\ref{Fargionfig5}.



\section{The early Light Nuclei photo-dissociation and their 20 EeV fragments}

\begin{figure}[t]
\begin{center}
\includegraphics[width=0.98\textwidth]{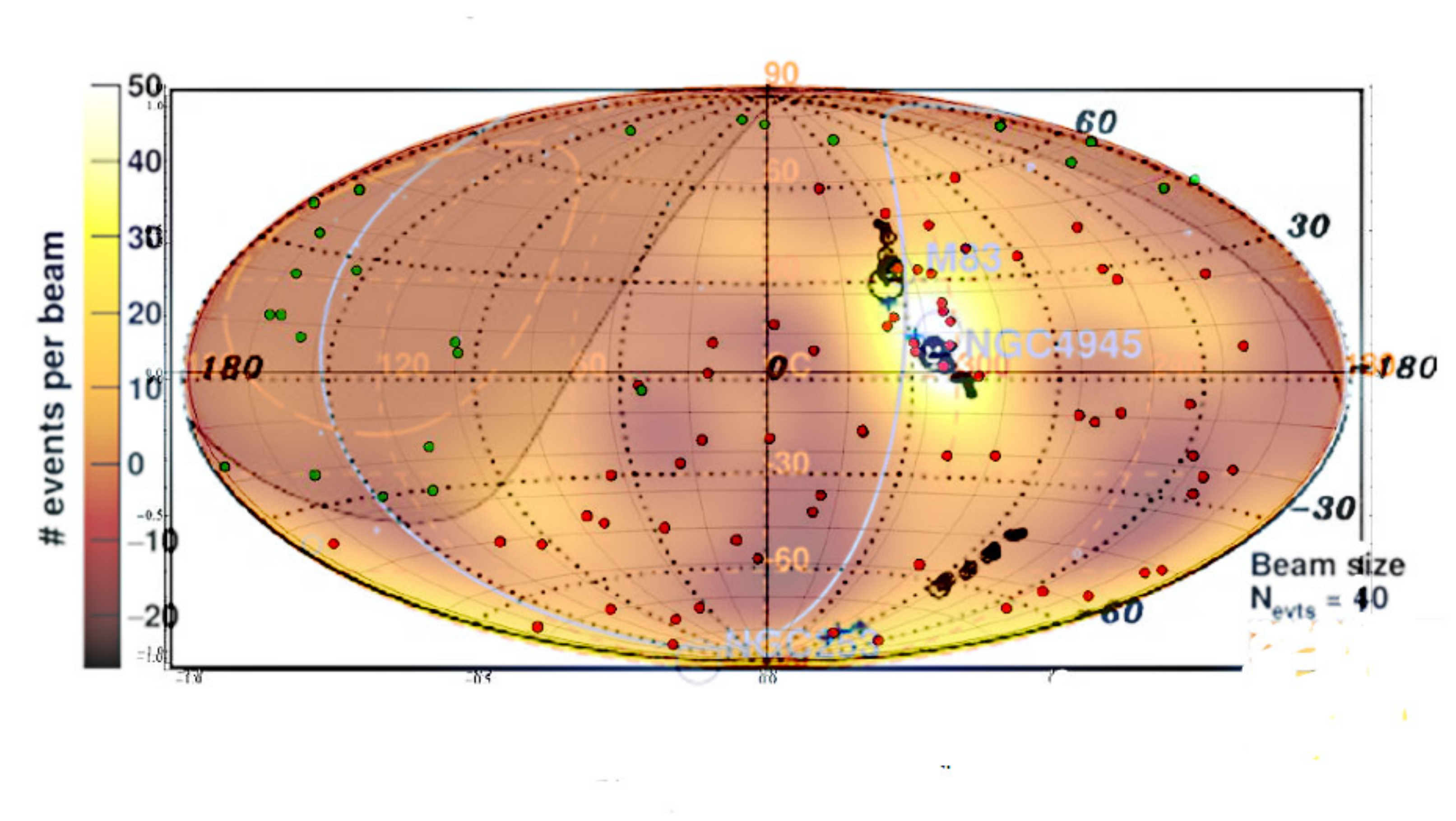}
\caption{This galactic UHECR map with in background the recent $39$ EeV anisotropy map, shows the earliest (2011) UHECR events and the 20 EeV smeared multiplet, up-down respect to the Cen~A position. We interpret these vertical spread of events as due to the planar galactic spiral magnetic fields, whose spin flip during the UHECR flight, leading to an up-down deflection of the charged He-like UHECR.
The one and a half deflection of the 20 EeV multiplet of the fragment events (respect harder ones) is larger than the $6 \cdot 10^{19}$ eV highest energy clustered events around Cen A, because their energy is nearly a third of the highest ones (He-like), but their charge is nearly half (Deuterium like) \cite{fargion2011coherent}.}
\label{Fargionfig6}
\end{center}
\end{figure}

We have suggested \cite{Fargion(2009b)} that these fragments had to arise and be detectable as multiplet events (from half to a fourth of the GZK energies).  Indeed they have possibly been observed by Auger in the 2011 at 20 EeV.
In the following map, see Fig.~\ref{Fargionfig6}, the Auger results from 2018 regarding the anisotropy of CR with E > 39 EeV \cite{PAO(2018)} are the background of these interesting narrow multiplets of clustering, as explained in the caption of Fig.~\ref{Fargionfig6}.

Their clustering in a multiplet may be telling us that the deflection is ruled, as expected, by the random galactic spiral alternate magnetic field lines.
Their size is well explained assuming a characteristic third of energy and half of a charge for these UHECR fragments (Deuterium).

\section{Names and locations of most UHECR clusterings}
\begin{figure}[t]
\begin{center}
\includegraphics[width=0.98\textwidth]{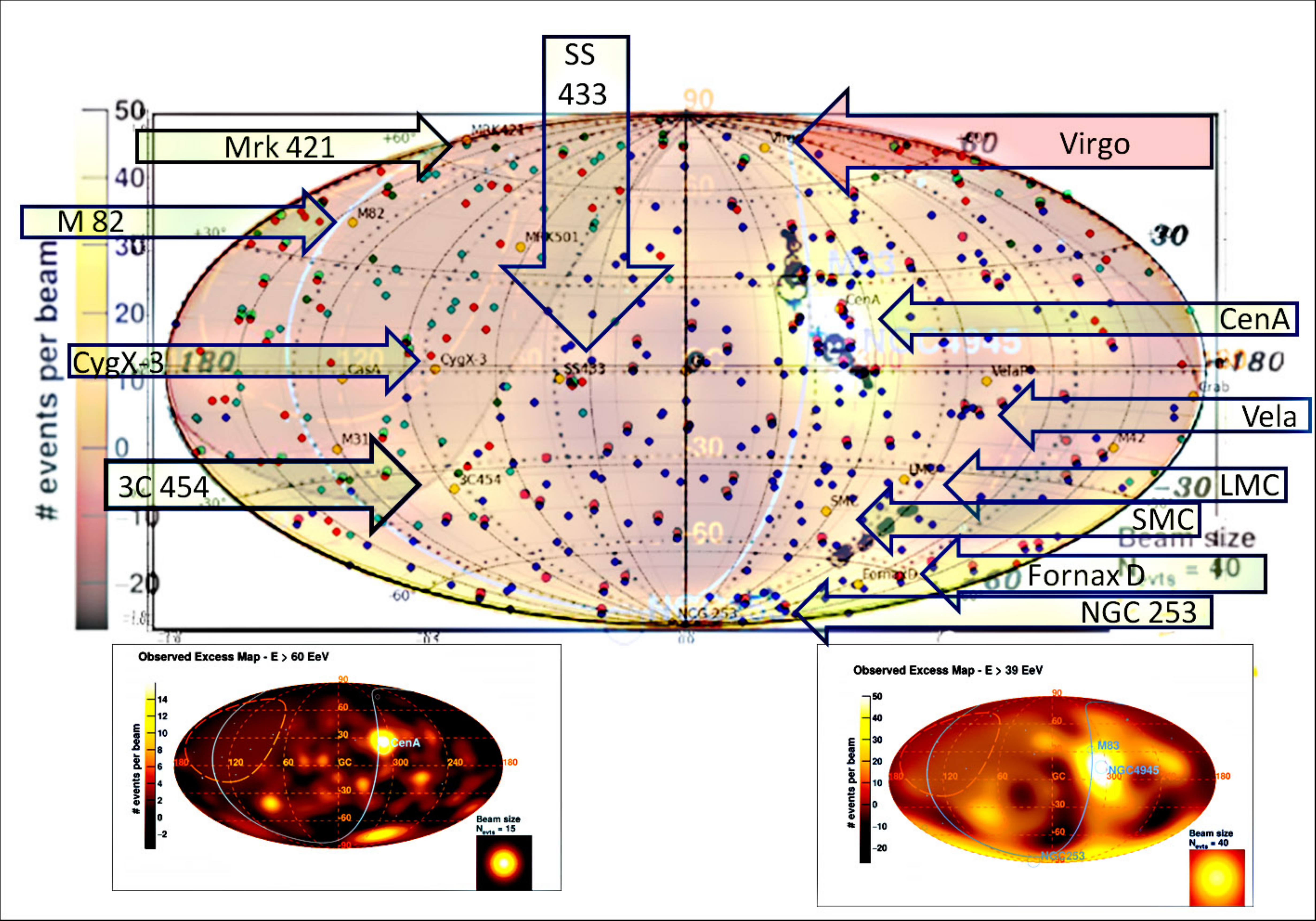}
\caption{In present galactic coordinate sky map we recalled and tagged all the main probable sources (or missing ones, as Virgo)\cite{Fargion2015Meaning}.
 A 39 EeV anisotropy map is in background.
 These candidate sources  are not different from previous ones discussed since 2008-2015. In particular we like to
 remind the very growing signals at Vela nearby sky, the weak LMC-SMC  clustering, the enhanced NGC~253 and Fornax~D area of events.
 In addition to the 3C~454, also far Mrk~421 and possibly the far AGN PKS 0208-512 might also play a role as sources of UHECR clustering; their UHE ZeV neutrino courier, hitting
 the relic cosmic neutrino background with mass (possibly at $\simeq 1.6 eV$ as the candidate sterile one) are leading to nucleon and antinucleon secondaries observed as UHECR cluster. }
\label{Fargionfig8}
\end{center}
\end{figure}

Now we consider the last individual  events over the last averaged anisotropy UHECR for highest $60$ EeV maps \cite{PAO(2018)}, (see Fig.\ref{Fargionfig8} small map in left corner).
In this higher energy the UHECR clusterings are more narrow and clearer, as we should expect from the deflection law.
There are  growing evidences  that favour Cen~A AGN as a main source  while NGC~253 signal appears to be affirming itself also as a good source candidate  in Auger sky. In the TA north sky M82 it is a possible, although off axis source.
There are weaker clusterings in different areas.
There it is a most relevant one, below the Vela source, and a very peculiar clustering
toward the far AGN 3C~454, see Fig.\ref{Fargionfig8}.

The very recent Auger clustering at $60$ EeV \cite{PAO(2018)} is shown in an averaged color map, see in left corner of the Fig.~\ref{Fargionfig8}.
Here we show the most certain (NGC~253, Cen~A, M82) as well as the most probable galactic or extragalactic names and locations and galactic coordinates.

\section*{Conclusion and Summary}
\subsection{Lightest nuclei UHECR smeared astronomy: Names and Locations}
The possible conclusion in the present short review of two decades of UHECR rush it is that we are finally seeing in the fog a first sky of UHECR sources. Their Light nuclei main signature guarantees two important results: Virgo UHECR cluster are obscured (as being Virgo  absent), the UHECR are in a very restricted local universe of just a few Mpc. This facilitated our identification of the UHECR possible locations tuned with the growing narrow scale anisotropy.
\subsection{10 EeV Dipole origination}
The 10 EeV dipole anisotropy cannot be (realistically) ruled by an extragalactic origin. Indeed if it were made by proton reaching us from cosmic Giga-Parsec volumes, their smeared tracks would trace Virgo (or even the Great Attractor).
However, the Auger UHECR maps at 10 EeV have just shown the dipole in an opposite direction of the sky where nearest galactic Vela, near  LMC, NGC~253 and Crab pulsar are shining and are located, see Figs.~\ref{Fargionfig3},\ref{Fargionfig4}.
They may be the responsible ones.
Moreover, if this Dipole anisotropy was due to UHECR protons (contrary to Auger light nuclei composition results, see Fig.\ref{Fargionfig5}) it would encompass huge cosmic volumes (several hundreds Mpcs), leading to a very averaged and smooth map, thus unable to reach the $7\%$ anisotropy but maybe as large as the cosmic dipole one, nearly $35$ times smaller. In the latter case anyway the anisotropy it would be mostly clustered in a Super-galactic plane or along a Great Attractor in contrast with the observed anisotropy size and arrival direction.

\subsection{Nearby Cen A, NGC~253, M82  sharp growing signals}
Cen A, the brightest and nearest radio AGN it is loudly found and confirmed in two decades.
 The mirror companion at the South galactic pole, NGC~253, is also blowing up at South galactic pole as well as the M82 AGN, might be the main source in the North galactic pole \cite{2015arXiv151208794F}.
 One may wonder why Cen A sits at the center of the UHECR celestial coordinate map and at the center of its South galactic Hot Spot,  while NGC~253 and M82 seem to be at the edges of their  Hot Spot, see Fig.\ref{Fargionfig8}.

A possible explanation it is that these two sources, coming from the North and South pole, did not suffer of a random deflection (as Cen A in near galactic spiral plane). Indeed they may have been bent by a main coherent magnetic field (a bubble magnetic skin on the galactic top of North and South poles) that are deflecting the charged UHECR spot mainly on one side of their optical direct gamma AGN position.
Future additional data might better confirm this solution; a weaker UHECR clustering tail might be  connecting the UHECR main spot to its optical gamma location. More data and composition signature are necessary, of course, to be sure that this it is the correct interpretation.

\subsection{Conclusions: The 3C~454, Z-Burst model and the cosmic relic sterile neutrino?}
Finally a main cornerstone for an exciting new physics might be linked with the 3C~454 AGN clustering of UHECR events (see Fig.\ref{Fargionfig8} in this paper and also Fig. 5, arrow A in \cite{Fargion2017NarrowClustering}).
Also Mrk~421 and PKS~0208-512  may play such a role in additional two hot spots.
Such a far source (3C~454) is near $2.5$ Giga-parsec distance, almost 50 times above the most penetrating proton UHECR (and almost thousand times the GZK He survival distance):
there is no way to reach directly to us overcoming the exponential ($\simeq e^{-50}$) GZK cutoff.
Therefore, the possible role of a relic mass neutrino of nearly one eV mass may play the role of a calorimeter able to capture a different courier (a ZeV anti-neutrino) in order to explain this puzzling connection \cite{Fargion.et.al.(1999)}, \cite{Fargion2007Splitting}.
More in detail the AGN may shine UHECR also at ZeV energies for neutrinos that cross the Universe with negligible absorption  and hit in a wide relic neutrino mass halo (several or tens Mpc wide) making a Z boson:
such forming UHE Z resonance, its decay in hadrons (involving nucleon-antinucleon pairs) is the final UHECR message to us.

The consequent presence of UHE EeV and ZeV tau neutrinos may also lead to detectable upward tau air-showers~\cite{Fargion(2000), Fargion.et.al.(2004), PAO(2009), PAO(2008)}.
The very recent and popular sterile neutrino at $1.6$ eV mass is incidentally well tuned to produce UHECR at a peak of $5 \cdot 10^{19}$ eV \cite{Fargion2007Splitting}. The thermalization of sterile neutrino is possible, if it does not influence much present experimental bounds, involving the constraints from the measured Z boson width and neutrino oscillations, as well as limits on the dark radiation effects in BBN, CMB and LSS.
The huge neutrino relic halo with a size as large as the characteristic GZK one (40-80 Mpc), may be the volume where a Z resonance may shower its secondary nucleon or antinucleon to us observed as common UHECR.

The road to disentangle such a possibility, quite different from light nuclei UHECR, might be a "personal tuned" test of the UHECR composition study (by its $X_{max}$ or slant depth, see \cite{Gaisser1977Reliability} ) study for each of the arrival direction source: CenA, M82, NGC~253 (light UHECR  nuclei events) versus Mrk421, 3C~454 (UHECR  nucleon and anti-nucleon signals) on the other side.

In a short summary, UHECR are very probably formed in AGN quasars and in micro-quasars (star forming regions),
powered by Black Hole binary tidal disruptions, both big and small but ejecting by directional jets, shining at different sizes, times and space.
The hadronic heavy nuclei jets are possibly fragmented and mostly survived far away only by the lightest nuclei able to reach us.
The nearest micro-quasars or early GRB \cite{Fargion(1999)} (as SS433 and maybe Vela and Crab later on gamma pulsar) may play a complementary  role, with LMC and SMC, at the "low 10 EeV" UHECR anisotropy, while highest UHECR are ruled by NGC253, M82, Cen~A, the nearest starburst or AGN in our Universe.

Few rare non galactic clusterings as the one around 3C~454 might be originated by the UHE ZeV neutrino scattering on relic cosmic ones, with a (originally sterile?) eV size mass.  Time and growing data as ad hoc test of the UHECR composition (the ones from 3C~454  should be nucleon like and not nuclei like), they will confirm or will wash away this exciting, extreme and fine tuned astro-particle solution.

\section*{Acknowledgements}
The work by MK was supported by Russian Science Foundation
and fulfilled in the framework of MEPhI Academic Excellence Project (contract
02.a03.21.0005, 27.08.2013).

\bibliographystyle{JHEP}
\bibliography{UHECR2018-bib-database}

\providecommand{\href}[2]{#2}\begingroup\raggedright\begin{thebibliography}{10}

\bibitem{AGASA(1998)}
M.~a. Takeda, N.~Hayashida, K.~Honda, N.~Inoue, K.~Kadota, F.~Kakimoto et~al.,
  \emph{Extension of the cosmic-ray energy spectrum beyond the predicted
  greisen-zatsepin-kuz'min cutoff}, {\emph{Physical Review Letters} {\bfseries
  81} (1998) 1163}.

\bibitem{GZK(1966)}
K.~Greisen, \emph{End to the cosmic-ray spectrum?},
  \href{https://doi.org/10.1103/PhysRevLett.16.748}{\emph{Physical Review
  Letters} {\bfseries 16} (1966) 748}.

\bibitem{GZK(1966)b}
G.~T. {Zatsepin} and V.~A. {Kuz'min}, \emph{{Upper Limit of the Spectrum of
  Cosmic Rays}}, {\emph{Soviet Journal of Experimental and Theoretical Physics
  Letters} {\bfseries 4} (1966) 78}.

\bibitem{PAO(2007)}
J.~Abraham, P.~Abreu, M.~Aglietta, C.~Aguirre, D.~Allard, I.~Allekotte et~al.,
  \emph{Correlation of the highest-energy cosmic rays with nearby extragalactic
  objects}, {\emph{Science} {\bfseries 318} (2007) 938}.

\bibitem{PAO(2017b)}
A.~Aab, P.~Abreu, M.~Aglietta, I.~Al~Samarai, I.~Albuquerque, I.~Allekotte
  et~al., \emph{Combined fit of spectrum and composition data as measured by
  the pierre auger observatory}, {\emph{Journal of Cosmology and Astroparticle
  Physics} {\bfseries 2017} (2017) 038}.

\bibitem{Fargion(2009b)}
D.~Fargion, \emph{Coherent and random uhecr spectroscopy of lightest nuclei
  along cen a: Shadows on gzk tau neutrinos spread in a near sky and time},  in
  \emph{Nuclear Instruments and Methods in Physics Research Section A, NIMA\/}
  \cite{fargion2011coherent}, 111.

\bibitem{PAO(2017)}
P.~A. Collaboration et~al., \emph{Observation of a large-scale anisotropy in
  the arrival directions of cosmic rays above 8 $10^{18} ev$}, {\emph{Science}
  {\bfseries 357} (2017) 1266}.

\bibitem{PAO(2011)}
P.~Abreu, M.~Aglietta, E.~Ahn, I.~F. d.~M. Albuquerque, D.~Allard, I.~Allekotte
  et~al., \emph{Search for signatures of magnetically-induced alignment in the
  arrival directions measured by the pierre auger observatory},
  {\emph{Astroparticle Physics} {\bfseries 35} (2012) 354}.

\bibitem{Fargion.2018}
D.~Fargion, P.~Oliva and P.~G. De~Sanctis~Lucentini, \emph{Uncorrelated far
  active galactic nuclei flaring with their delayed ultra high energy cosmic
  rays events}, {\emph{JPS Conf. Proc. 19, 011010 (2018),(UHECR2016)} }.

\bibitem{Fargion.et.al.(1999)}
D.~Fargion, B.~Mele and A.~Salis, \emph{Ultra-high-energy neutrino scattering
  onto relic light neutrinos in the galactic halo as a possible source of the
  highest energy extragalactic cosmic rays}, {\emph{The Astrophysical Journal}
  {\bfseries 517} (1999) 725}.

\bibitem{Fargion2007Splitting}
D.~Fargion, D.~D'Armiento, O.~Lanciano, P.~Oliva, M.~Iacobelli,
  P.~De~Sanctis~Lucentini et~al., \emph{Splitting neutrino masses and showering
  into sky},
  \href{https://doi.org/10.1016/j.nuclphysbps.2007.02.090}{\emph{Nuclear
  Physics B - Proceedings Supplements} {\bfseries 168} (2007) 292}.

\bibitem{Fargion2004Colaiuda}
D.~Fargion and A.~Colaiuda, \emph{Gamma rays precursors and afterglows
  surrounding uhecr events: Z-burst model is still alive}, {\emph{Nuclear
  Physics B-Proceedings Supplements} {\bfseries 136} (2004) 256}.

\bibitem{Fargion2017NarrowClustering}
D.~Fargion, P.~Oliva, P.~G. D.~S. Lucentini, D.~D'Armiento and P.~Paggi,
  \emph{Uhecr narrow clustering correlating icecube through-going muons},
  \href{https://doi.org/https://doi.org/10.1016/j.nuclphysbps.2017.06.036}{\emph{Nuclear
  and Particle Physics Proceedings} {\bfseries 291-293} (2017) 195 }.

\bibitem{Fargion(2000)}
D.~Fargion, \emph{Discovering ultra-high-energy neutrinos through horizontal
  and upward $\tau$ air showers: evidence in terrestrial gamma flashes?},
  {\emph{The Astrophysical Journal} {\bfseries 570} (2002) 909}.

\bibitem{Fargion.et.al.(2004)}
D.~Fargion, P.~D.~S. Lucentini, M.~De~Santis and M.~Grossi, \emph{Tau air
  showers from earth}, {\emph{The Astrophysical Journal} {\bfseries 613} (2004)
  1285}.

\bibitem{PAO(2009)}
J.~Abraham, P.~Abreu, M.~Aglietta, C.~Aguirre, E.~Ahn, D.~Allard et~al.,
  \emph{Limit on the diffuse flux of ultrahigh energy tau neutrinos with the
  surface detector of the pierre auger observatory}, {\emph{Physical Review D}
  {\bfseries 79} (2009) 102001}.

\bibitem{PAO(2008)}
J.~Abraham, P.~Abreu, M.~Aglietta, C.~Aguirre, D.~Allard, I.~Allekotte et~al.,
  \emph{Upper limit on the diffuse flux of ultrahigh energy tau neutrinos from
  the pierre auger observatory}, {\emph{Physical Review Letters} {\bfseries
  100} (2008) 211101}.

\bibitem{Huchra2012RedShift}
J.~P. {Huchra}, L.~M. {Macri}, K.~L. {Masters}, T.~H. {Jarrett}, P.~{Berlind},
  M.~{Calkins} et~al., \emph{{The 2MASS Redshift Survey-Description and Data
  Release}}, \href{https://doi.org/10.1088/0067-0049/199/2/26}{\emph{\apjs}
  {\bfseries 199} (2012) 26} [\href{https://arxiv.org/abs/1108.0669}{{\ttfamily
  1108.0669}}].

\bibitem{Fargion(2008)}
D.~Fargion, \emph{Light nuclei solving the auger puzzles: the cen-a imprint},
  {\emph{Physica Scripta} {\bfseries 78} (2008) 045901}.

\bibitem{Fargion(2009)}
D.~Fargion, D.~D'Armiento, P.~Paggi and S.~Patri', \emph{Lightest nuclei in
  uhecr versus tau neutrino astronomy},
  \href{https://doi.org/https://doi.org/10.1016/j.nuclphysbps.2009.03.083}{\emph{Nuclear
  Physics B - Proceedings Supplements} {\bfseries 190} (2009) 162 }.

\bibitem{Fargion2015Meaning}
D.~Fargion, G.~Ucci, P.~Oliva and P.~G. De~Sanctis~Lucentini, \emph{The meaning
  of the uhecr hot spots-a light nuclei nearby astronomy}, {\emph{EPJ Web Conf}
  {\bfseries 99} (2015) 08002}.

\bibitem{Kampert(2016)}
K.-H. Kampert and P.~A. Collaboration, \emph{Ultra-high energy cosmic rays:
  Recent results and future plans of auger}, {\emph{AIP Conference Proceedings
  1852, 040001 (2017)} }.

\bibitem{Deligny(2017)}
O.~Deligny, K.~Kawata and P.~Tinyakov, \emph{Measurement of anisotropy and
  search for uhecr sources}, {\emph{Progress of Theoretical and Experimental
  Physics, Volume 2017, Issue 12, 1 December 2017, 12A104,,arXiv:1702.07209}
  (2017) }.

\bibitem{Joshi(2018)}
J.~Joshi, M.~Miranda, L.~Yang and S.~Razzaque, \emph{Very high-energy gamma-ray
  signature of ultrahigh-energy cosmic-ray acceleration in centaurus a},
  {\emph{arXiv:1804.06093} (2017) }.

\bibitem{2015arXiv151208794F}
D.~Fargion, P.~Oliva and G.~Ucci, \emph{Why not any tau double bang in icecube,
  yet?}, {\emph{PoS FRAPWS2014 (2016) 028, arXiv:1512.08794} (2015) }.

\bibitem{PAO(2017c)}
M.~Boh{\'a}{\v{c}}ov{\'a}, P.~A. Collaboration et~al., \emph{Highlights from
  the pierre auger observatory}, {\emph{Nuclear and Particle Physics
  Proceedings} {\bfseries 291} (2017) 82}.

\bibitem{PAO(2018)}
A.~Aab, P.~Abreu, M.~Aglietta, I.~Albuquerque, I.~Allekotte, A.~Almela et~al.,
  \emph{Indication of anisotropy in arrival directions of ultra-high-energy
  cosmic rays through comparison to the flux pattern of extragalactic gamma-ray
  sources}, {\emph{arXiv preprint arXiv:1801.06160} (2018) }.

\bibitem{fargion2011coherent}
D.~Fargion, \emph{Coherent and random uhecr spectroscopy of lightest nuclei
  along cen a: Shadows on gzk tau neutrinos spread in a near sky and time},
  {\emph{Nuclear Instruments and Methods in Physics Research Section A, NIMA}
  {\bfseries 630} (2011) 111}.

\bibitem{Gaisser1977Reliability}
T.~K. {Gaisser} and A.~M. {Hillas}, \emph{{Reliability of the method of
  constant intensity cuts for reconstructing the average development of
  vertical showers}}, {\emph{International Cosmic Ray Conference} {\bfseries 8}
  (1977) 353}.

\bibitem{Fargion(1999)}
D.~Fargion, \emph{On the nature of grb-sgrs blazing jets}, {\emph{Astronomy and
  Astrophysics Supplement Series} {\bfseries 138} (1999) 507}.

\end{thebibliography}\endgroup
\end{document}